\documentstyle[epsf,aps,preprint,eqsecnum]{revtex}

\begin{document}

\bibliographystyle{prsty}

\title{
Quantum-classical phase transition of the escape rate in a biaxial
           spin system with an arbitrarily directed magnetic field
}

\author{Chang-Soo Park}
\address{Department of Physics,  Dankook University,  Cheonan 330-714,
         Korea\\}
\author{Sahng-Kyoon Yoo}
\address{Department of Physics, Seonam University, Namwon, Chunbuk 590-711,
         Korea\\}
\author{Dal-Ho Yoon}
\address{Department of  Physics, Chongju University, Chongju 360-764,
         Korea\\
\smallskip
{\rm (\today)}
\bigskip\\
\parbox{14.2 cm}
{\rm
We investigate the escape rate of a biaxial spin particle with an
arbitrarily directed magnetic field in the easy plane, described by
Hamiltonian ${\cal H}= -AS_z^2 - BS_x^2 - H_x S_x - H_z S_z , \;
(A > B > 0)$. We derive an effective particle potential by using
the method of particle mapping. With the help of the criterion for
the presence of a first-order quantum-classical transition of the escape rate
we obtained various phase boundary curves depending on the anisotropy
parameter $b \equiv B/A$ and the field parameters $\alpha_{x, z}
\equiv H_{x, z}/AS$: $\alpha_{zc}(b_c)$'s, $\alpha_{xc}(b_c)$'s, and
$\alpha_{zc} = \alpha_{zc}(\alpha_{xc} )$. It is found from $\alpha_{zc}(b_c)$'s
and $\alpha_{xc}(b_c)$'s
that the first-order region decreases as $b$ and $\alpha_x$ (or $\alpha_z$) increase.
The phase boundary line $\alpha_{zc} = \alpha_{zc} (\alpha_{xc} )$ shows
that compared with the uniaxial system, both the first- and
second-order regions are diminished due to the transverse
anisotropy.
Moreover, it is observed that, in the limit $\alpha_{xc} \rightarrow
0$, $\alpha_{zc}$ does not coincides with the coercive field line,
which yields more reduction in the first-order region.
We have also computed the crossover
temperatures at the phase boundary : $T_c(b_c)$, $T_c(\alpha_{xc} , \alpha_{zc})$.
\smallskip
\begin{flushleft}
PACS number (s) :75.45.+j, 75.50.Tt
\end{flushleft}
}
}

\maketitle

\section{INTRODUCTION}
\label{into}

Recently there have been intensive studies on the
quantum-classical phase transition of the escape rate in a single
domain magnetic particle with many spins. In such a particle the magnetization direction of
a collection of spins is oriented such that the magnetocrystalline
anisotropy energy is at a stable or metastable minimum depending
on the existence of an external magnetic field. The escape from a
stable or metastable state is governed by classical thermal
activation rate, which is proportional to $\exp(-E_0 /k_B T)$,\cite{ston} at
high temperature, and by quantum tunneling at temperatures below
the energy barrier $E_0$. When these two escape rates are same
there exists a crossover temperature $T_0$ at which the transition
between classical and quantum regimes occurs. The transition can
be either first-order or second-order. In the first-order
transition the escape rate abruptly changes from the
temperature-dependent thermal activation process to a practically
temperature-independent quantum tunneling, so that the first
derivative of the escape rate at the crossover temperature changes
discontinuously. In the second-order transition, however, the
escape rate changes smoothly from classical regime to
temperature-dependent quantum tunneling (thermally assisted
tunneling), and has a discontinuity of the second derivative.
The determination of the transition order is closely related to
the shape of the potential barrier which is controlled by the
anisotropy constant and the external magnetic field.

The quantum-classical transition of the escape rate was
investigated by Affleck \cite{affleck} and Larkin and Ovchinnikov.
\cite{larkin} By using instanton technique they demonstrated that
under certain assumption on the shape of potential barrier a
smooth interpolation between the periods of oscillations at the
bottom and the top of Euclidean potential well can be made, which leads to
a second-order phase transition between classical and quantum
regimes. Later, Chudnovsky \cite{chud} observed that the order of
the phase transition in the crossover from classical activation
to thermally assisted tunneling completely depends on the shape of
the potential barrier. He has shown that the behavior of the
energy-dependent period of oscillations $\tau (E)$ in Euclidean
potential determines the order of the quantum-classical
transition; if the Euclidean period increases monotonically with
the increasing energy $E$ from the bottom of the Euclidean
potential the transition is second-order, but if $\tau(E)$ is
nonmonotonic such that it has a minimum at $E=E_1$ which is
smaller than the potential barrier $E_0$ the first-order
transition occurs. More recently, Gorokhov and Blatter
\cite{blatter} have obtained a sufficient condition for the
first-order quantum-classical phase transition by looking at the
behavior of the oscillation period in Euclidean time as a function
of oscillation amplitude near the barrier top in the
two-dimensional string model. In this case the first-order
transition appears when the amplitude-dependent period $\tau(a)$,
where $a$ is the amplitude, is smaller than the zero amplitude
period $\tau(0)$ near the
barrier top. This method has been subsequently extended to a
quantum-mechanical model where mass has coordinate-dependence.
\cite{dkpark}

The above approaches have been applied to the quantum-classical phase
transition of the escape rate in a single domain spin system. Up
to now two types of spin systems have been studied intensively:
uniaxial and biaxial systems. For the uniaxial system, such as
high-spin molecular magnet $\rm{Mn_{12} Ac}$,\cite{fried} two
models have been considered: one with a transverse field and the
other with an arbitrarily directed field, described by
Hamiltonians ${\cal{H}}=-DS_z^2 - H_x S_x$ \cite{gara} and
${\cal{H}}=-DS_z^2 - H_x S_x - H_z S_z$,\cite{marti}
respectively. The biaxial spin system, such as iron
cluster $\rm{Fe_8}$,\cite{ohm} has attracted more attention, and
several models have been taken into account. Liang $\it{et \; al.}$\cite{liang} considered
a model without an applied field, ${\cal{H}}=K(S_z^2 + \lambda S_y^2 ), (0<\lambda<1)$
by using the periodic instanton approach and demonstrated that
the coordinate-dependent effective mass plays an important role
for the presence of the first-order transition. Based on the same approach
Lee $\it{et \: al.}$ \cite{lee} investigated the biaxial spin model with a
transverse field, ${\cal{H}}=K(S_z^2 + \lambda S_y^2 )-H_y S_y$,
and showed that the nonconstant mass which depends on both coordinate
and field is important for the occurrence of the first-order transition.
However, this approach involves a restricted
range of applicability of $\lambda << 1$ for the biaxial spin
system with a field. Such a restriction can be avoided by
introducing the method of particle mapping.\cite{zas} The same model has been considered
by Kim \cite{kim} who used a quasiclassical method based on the
particle mapping and found an analytical
form of the phase boundary curve between first- and second-order
transitions. The phase boundary of a biaxial system with longitudinal field
described by ${\cal{H}}=K(S_z^2 + \lambda S_y^2 ) - H_x S_x$ has also been
obtained by Garanin and Chudnovsky \cite{garanin} employing a perturbation
approach with respect to the transverse anisotropy, by Kim \cite{kim}
using the quasiclassical method based on particle mapping, and by
Park $\it{et \; al.}$ \cite{cspa} with the help of the Gorokhov and Blatter's
criterion \cite{blatter} and the particle mapping.

In this paper we study the phase transition of the escape rate of a biaxial
spin system with an arbitrarily directed magnetic field in the
easy plane. In this case the transverse and longitudinal fields
coexist, and there are three parameters which can be controlled by
experiment: the anisotropy constant and two field parameters.
In order to find a phase diagram for the transition orders these should be considered
simultaneously.
We will use the method of particle mapping to derive an effective
particle potential from the spin Hamiltonian. Then, by applying the
criterion developed in Refs. \onlinecite{blatter,dkpark} to this
potential we will find various phase diagrams depending on the
three parameters. Especially, we will obtain a phase boundary
curve between the first- and the second-order transitions
for the iron cluster $\rm Fe_8$ in which a relation between the transverse
and longitudinal field parameters is given. We will also find
the crossover temperature at the phase boundary.

In the following section, we present a derivation of the effective
particle potential based on the method of particle mapping.  Here, we will
take a constant mass so that the
effective potential includes all the coordinate dependence. We will also
review the sufficient condition for the presence of the first-order transition. In
Sec. III we show various phase boundary curves between the
first- and second-order transitions. For completeness we will also
briefly discuss two special cases: the
biaxial systems with the transverse field only
and with the longitudinal field only. These diagrams are compared
with the previously obtained results in uniaxial system. The
crossover temperature at the phase boundary is also computed here.
Finally, there will be a summary and discussions in Section IV.

\section{PARTICLE MAPPING AND THE CRITERION FOR THE FIRST-ORDER TRANSITION}
\label{map}
Consider a biaxial single domain magnetic particle with $XOZ$ easy-plane and the easy
$Z$-axis in the $XZ$ plane. When an external field is applied
along an arbitrary direction in the $XZ$ plane the Hamiltonian can
be described by
\begin{equation}
{\cal H}=-AS_z^2 - BS_x^2 - H_x S_x - H_z S_z ,
\end{equation}
where $A$ and $B$ are the longitudinal and transverse anisotropy constants, respectively
satisfying $A>B>0$. Our model is equivalent to ${\cal H} = K(S_z^2 + \lambda S_y^2) - H_x S_x
-H_y S_y$ if we set $A=K$, $B=(1-\lambda)K$.
For convenience, we introduce dimensionless transverse anisotropy parameter $b \equiv B/A (<1)$
and field parameters $\alpha_x \equiv H_x /SA$, $\alpha_z \equiv H_z /SA$
where $S$ is the spin number. Following Ref. \onlinecite{zas} this
spin problem can be reduced to a particle moving in an
effective potential. The equivalent particle Hamiltonian can be
written as
\begin{equation}
{\cal H} = -\frac{1}{2m} \frac{d^2}{dx^2} + V(x),
\end{equation}
where $m \equiv 1/2A$ is the particle mass, and $V(x) \equiv A v(x)$
is the effective particle potential given by
\begin{eqnarray}
v(x) = \frac{1}{4{\rm dn}^2 x}[&S^2 (\alpha_x {\rm sn}x
- \alpha_z {\rm cn}x)^2 - 4bS(S+1) \nonumber\\
&-2S(2S+1)(b\alpha_z {\rm sn}x +\alpha_x {\rm cn}x )
]
\end{eqnarray}
in which ${\rm sn}x$, ${\rm cn}x$, ${\rm dn}x$ are the Jacobian
elliptic functions with modulus $k^2 = 1 - b$. The
Schr\"{o}dinger-like equation corresponding to this Hamiltonian is
${\cal H}\Psi (x) = E\Psi (x)$ where $\Psi(x)$ is the particle
wave function given by
\begin{eqnarray}
\Psi(x) &=&\left(\frac{{\rm cn}x}{{\rm dn}x} \right)^S \exp \left[
\frac{\alpha_z S}{2\sqrt{1-b}} \tanh^{-1} (\sqrt{1-b} {\rm sn}x)
\right]\nonumber\\
&\times& \exp \left[-\frac{\alpha_x S}{2\sqrt{b(1-b)}}\tan^{-1}
\left( \sqrt{\frac{b}{1-b}}\frac{1}{{\rm cn}x} \right) \right]
\Phi(x)\nonumber\\
\end{eqnarray}
with
\begin{equation}
\Phi(x) = \sum_{\sigma = -S}^{S}
\frac{C_{\sigma}}{\sqrt{(S+\sigma)! (S-\sigma)!}} \left(
\frac{{\rm sn}x +1}{{\rm cn}x} \right)^{\sigma} .
\end{equation}
In Fig.1 the effective particle potential is drawn. It has a
metastable minimum with asymmetric barriers: small and large ones.
The heights and widths of these barriers are governed by the
anisotropy and field parameters $b,\;\alpha_x ,\;\alpha_z$. For a
given value of $b$ the height of the small barrier decreases as $\alpha_x$
or $\alpha_z$ increases and vanishes at a critical value, while that of the
large barrier increases. The critical value is determined by the metastability
condition which, for the present model, is derived as\cite{meta}
\begin{equation}
\alpha_{xm}^{2/3} + \alpha_{zm}^{2/3} =
[2(1-b)]^{2/3}.
\end{equation}
For a given value of $b$ the metastable state exists inside the region closed by
the curve $\alpha_{xm}=\alpha_{xm}(\alpha_{zm})$.
We note here that because of the transverse anisotropy
the metastability region of the biaxial system is smaller than the
uniaxial system which corresponds to the case of
$b=0$.\cite{compare1}

The escape rate at temperatures below the barrier height $E_0$ can be obtained by taking
thermal average over tunneling probabilities. In the semiclassical approximation
this can be expressed as
\begin{equation}
\Gamma \propto \int_0^{E_0} dE P(E)e^{-E/T},
\end{equation}
where the metastable minimum is chosen to be zero energy. The
tunneling probability $P(E)$ can be approximated by the WKB
exponent, $P(E) \sim e^{-S(E)}$ in which $S(E)$ is the Euclidean
action defined by
\begin{equation}
S(E) = 2\sqrt{2m} \int_{x_1 (E)}^{x_2 (E)} dx \sqrt{v(x) - E},
\end{equation}
where $x_1 (E), x_2 (E)$ are the turning points corresponding to
energy $E$. For large spin system the semiclassical approximation
is well applicable so that we can neglect the contributions from
paths fluctuating around the semiclassical saddle point path which
minimizes the Euclidean action $S(E)$. Within this approximation
the escape rate becomes
\begin{equation}
\Gamma \sim \exp \left[-\frac{S_{T\rm min}(E)}{T} \right] ,
\end{equation}
where $S_{T\rm min}(E)$ is the minimum of an energy function
\begin{equation}
S_T (E) = S(E) + E/T,
\end{equation}
which is called {\it thermon} action\cite{chud} or {\it periodic
instanton} action in field theory.
The condition for $S_T(E)$ to have minimum requires  $dS_T(E)/dE = 0$.
Since the first derivative of an action with respect to
energy in a potential well brings about the oscillation period we
can write
\begin{equation}
\tau (E) = - \frac{dS(E)}{dE} = \frac{1}{T},
\end{equation}
where negative sign attributes to the Euclidean potential in which
the energy has negative values, and hence
$\tau (E)$ is an Euclidean time oscillation period. Since the
Euclidean action $S(E)$ is zero at $E=E_0$ (i.e., at the top of the
barrier) the minimum of the energy function becomes
\begin{equation}
S_{T{\rm min}}(E_0 ) \equiv S_0 = \frac{E_0}{T}
\end{equation}
which is just the exponent of the thermal activation rate, i.e., the
{\it thermodynamic action}.\cite{chud}

The type of phase transition is determined by the behavior of $\tau
(E)$ with energy $E$. As has been well analyzed by Chudnovsky\cite{chud} a
first-order quantum-classical transition takes place when $\tau (E)$
has a minimum at some energy $E_1 \:( < E_0 )$. In this case the
crossover temperature $T_0 = 1/\tau(E_0 )$ is lower than $T_1 = 1/\tau(E_1
)$, and there exists a temperature $T_c \; (T_0 < T_c < T_1 )$
corresponding to an Euclidean oscillation period
$\tau_c$ at which the classical escape rate with $S_0$ and the
quantum escape rate with the minimum thermon action $S_{T\rm min}$
are connected, but their first derivatives with respect to $T$
are discontinuous. That is, the semiclassical saddle point
path jumps from the quantum regime to the classical one at temperature
$T_c$. It can be noted from the behavior of $\tau(E)$ that the first-order
transition exist if the Euclidean oscillation period at an energy near
the bottom of the Euclidean potential is smaller than $\tau(E_0)$,
the period corresponding to the crossover temperature.

The above argument can be generalized to the case in which the period is
a function of the oscillation amplitude $a$ near the bottom of the
Euclidean potential. When $E$ approaches to $E_0$ the dynamics in
the Euclidean potential well becomes small oscillations. At $E=E_0$
the oscillation amplitude is zero, and the thermon action becomes
the thermodynamic action. In this limit the
period can be related to the crossover temperature $T_0$ as
following:
\begin{equation}
T_0 = 1/\tau(0) = \omega_0 /2\pi,
\end{equation}
where $\tau(0)$ is a period corresponding to zero amplitude,
and $\omega_0$ is defined as
\begin{equation}
\omega_0 = \sqrt{-\frac{v''(x_0 )}{m}},
\end{equation}
where $x_0$ is the position of the top of potential
barrier. The solution of small oscillations near this point can be obtained from
the Euclidean Euler-Lagrange equation, and the oscillation period
of the solution can be
expressed by $\tau(a) = 2\pi / \omega$. From the above discussion the
condition for the presence of the  first-order quantum-classical
transition is then given by $\tau(a) < \tau(0)$. The difference $\tau(a) -
\tau(0)$ satisfying this condition can be calculated
by the perturbation method in which the oscillation amplitude $a$ is
used as a perturbation parameter.\cite{dkpark} For the case of constant mass
the condition becomes
\begin{equation}
- \frac{5}{24} \frac{v'''^2 (x_0 )}{v''(x_0 )} + \frac{1}{8}
v''''(x_0 ) \: < \: 0.
\end{equation}

Below we will use this criterion to compute the phase boundary
curve between the first- and the second-order transitions. Once we
obtain the boundary curve we can also calculate the crossover
temperature at the phase boundary from Eq.(13).

\section{PHASE BOUNDARY LINES BETWEEN THE FIRST- AND SECOND-ORDER TRANSITIONS}
\label{two}

We start this section by considering two special cases where either the
longitudinal field or the transverse field is applied, from which
we deduce some important features that can be directly applied to the case
with arbitrarily directed field. We first consider the model with transverse field only.
When the longitudinal field is zero the model can be described by
\begin{equation}
{\cal H} = -AS_z^2 -BS_x^2 - H_x S_x .
\end{equation}
This is equivalent to the previously studied models; if we set $A=K, \:
B=(1-\lambda)K$ it is same as the model considered in
Ref.\onlinecite{lee}, and if $A - B =K_{\parallel}, \: B = K_{\perp}$
it becomes the model of Ref.\onlinecite{kim}. The present
approach based on the criterion described above, however, is different from theirs.
In this case since $\alpha_z = 0$
in Eq.(3) the effective particle potential is reduced as
\begin{equation}
v_t (x) = \frac{S^2 \alpha_x^2 {\rm sn}^2 x
       -2S(2S+1)\alpha_x {\rm cn}x - 4bS(S+1)}{4{\rm dn}^2
       x},\nonumber\\
\end{equation}
and the corresponding wave function is given by
\begin{eqnarray}
\Psi_t (x) &=& \left(\frac{{\rm cn}x}{{\rm dn}x} \right)^S
\nonumber\\
          &\times& \exp \left[-\frac{\alpha_x S}{2\sqrt{b(1-b)}}\tan^{-1}
   \left(\sqrt{\frac{b}{1-b}}\frac{1}{{\rm cn}x} \right)
   \right]
\Phi(x). \nonumber\\
\end{eqnarray}

We note that the effective particle mass has no coordinate
dependence, which is different from previous models where the
coordinate dependent mass played a crucial role to present the
first-order transition.\cite{lee,zas} In converting the spin problem to a
particle one, whether the mass depends on the coordinate
is a matter of how to set up the Schr\"{o}dinger-like equation in
the process of particle mapping (see Appendix). In our case all
the coordinate dependence is included in the effective particle
potential. As we can see below our approach gives the same results
as Ref.\onlinecite{kim}.

The potential $v_t (x)$ has now small and large barriers with same
minima. Since the mass is independent of the coordinate it is obvious
that the escape over the small barrier dominates.\cite{compare2}
For large spin system, such as $S \sim S+1 \sim \tilde{S} \equiv
S+1/2$, the top of the small barrier is located at $x_0 = {\rm sn}^{-1}0 =
0$. By equating both sides of the criterion Eq.(15) and evaluating the
the derivatives of the potential $v_t (x) $ at $x = x_0$ we obtain an
equation of the phase boundary line between first- and
second-order transitions:
\begin{equation}
\alpha_{tc} (b_c ) = \frac{1-16b_c + 16b_c^2 +
\sqrt{1+32b_c -32b_c^2 } }{4(1-2b_c )}
\end{equation}
where the subscript $c$ represents that the anisotropy and field
parameters are the values taken at the phase
boundary. From Eq.(13), the transition temperature at the phase boundary $T_c$
is then obtained to be
\begin{equation}
\frac{T_{tc}}{\tilde{S}A} =
\frac{1}{2\pi}\sqrt{\frac{3\alpha_{tc}(b_c)}{2(1-2b_c )}},
\end{equation}
where $\alpha_{tc}(b_c )$ is given in Eq.(19). These results are
consistent with those in Ref.\onlinecite{kim} if we realize that
$\alpha_x = 2(1-b)h_x$, $b=k_t / (1+k_t )$.

The phase diagram for the case with longitudinal field only based
on the present approach has already been obtained by
us,\cite{cspa} so we just quote the results here:
\begin{eqnarray}
\alpha_{lc}(b_c ) = 2(1-b_c ) \sqrt{\frac{1-2b_c }{1+b_c }},
\nonumber\\
\frac{T_{lc}}{\tilde{S}A} = \frac{\sqrt{3}b_c }{\pi}
\sqrt{\frac{1-b_c}{1+b_c}}.
\end{eqnarray}

In Fig.2 we have plotted the phase boundary curves $\alpha_{tc}(b_c)$,
$\alpha_{lc}(b_c)$ and the coercive field line $\alpha_{tm}=\alpha_{lm}
\equiv 2(1-b)$. The first-order transition exist below the curves
$\alpha_{tc}(b_c)$ and $\alpha_{lc}(b_c)$.
In the region between these lines and the coercive field line the second-order
transition lies. From these results we observe some important features.
First, as the transverse anisotropy $b$ increases both
$\alpha_{tc}$ and $\alpha_{lc}$ decrease, and become zero at
$b=1/2$, i.e., the first-order region decreases in both cases.
Thus, the biaxial system has smaller first-order region than the
uniaxial system.
Second, for a given value of $b (< 1/2)$ the longitudinal field
case has larger first-order region than the transverse field case.
As we will see below these are common in the case with an
arbitrarily directed field.

The above features can be understood as follows. In the uniaxial system without
transverse field the spin operator commutes with the Hamiltonian,
and thus the spin becomes a constant of the motion (i.e., no
dynamics). Moreover, in this case the barrier of the effective
particle potential becomes infinitely thick so that there is no
tunneling, and hence no quantum-classical phase transition. In the
presence of the transverse field, however, the uniaxial spin
system becomes a dynamical one,
and so the tunneling occurs. In the limit of very small
transverse field the top of the effective potential barrier
becomes flat, which is favorable to the first-order transition.
When the transverse field increases, however, the situation
becomes unfavorable to the presence of the first-order. Now, in
the biaxial system the transverse anisotropy also gives dynamical
origin to the spin system, and thus enforces the escape process from
the metastable state, which leads to the suppression of the first-oder
transition. Therefore, the decrease of $\alpha_{lc}(b_c)$ is
caused by the transverse anisotropy only, whereas the boundary
line $\alpha_{tc}(b_c)$ is affected by both the transverse field and
transverse anisotropy.

We now consider the case with arbitrarily directed field. In this
case the three parameters $b, \; \alpha_x ,\; \alpha_z$ should be
treated simultaneously, which is not a simple problem. In the
present work we will fix one parameter and then compute the phase boundary
with the other two parameters. We first calculate the phase
boundary lines $\alpha_{zc}(b_c)$'s for several values of $\alpha_x$,
and $\alpha_{xc}(b_c)$'s for different values of $\alpha_z$,
which are shown in Fig.3. An immediate observation is that the
first-order region for a given $\alpha_x$ (or $\alpha_z$) diminishes as $b$
increases, which shows the same trend as the $\alpha_x = 0$ (or $\alpha_z =0$)
case.
We also note that the first-order region becomes smaller for
increasing values of $\alpha_x$'s (or $\alpha_z$'s). This can be readily
explained by the metastability condition in Eq.(6). When $\alpha_x$
and $b$ are given this condition yields the coercive field
line for $\alpha_z$ such as $\alpha_{zm}(\alpha_x , b)=2[(1-b)^{2/3}-
(\alpha_x/2)^{2/3}]^{3/2}$. From this it can be seen that the metastability
region, where the escape process can be considered, is shrunk on
the whole, which in turn suppresses the first-order transition
region.

Next, we compute the phase boundary for a fixed value of $b$. This
is the case when a specific sample is prepared in experiment.
Here, we take the molecular iron cluster $\rm{Fe}_8$ which has
been studied in the previous experiment.\cite{ohm} In this case
the transverse anisotropy parameter is given by
$b=0.29$. In Fig.4 we draw the phase boundary $\alpha_{zc}(\alpha_{xc})$
and the metastability condition line $\alpha_{zm}(\alpha_{xm})$
for $b=0.29$.
Since the metastability region of the biaxial system is smaller
than the unixail system we can see both the first- and
second-order regions become smaller than the uniaxial
system.\cite{marti} The point of intersection with
$\alpha_x$-axis is larger than that of $\alpha_{zc}(0)$, which
reveals that the case with longitudinal field only has larger
first-order region than the transverse field only case
.
An interesting feature in this diagram is that
unlike the uniaxial system $\alpha_{zc}$ does not coincides with the
coercive field $\alpha_{zm}$ at zero transverse field. It can thus
be realized that the first-order region is reduced more than the
second-order region. In fact this is anticipated if we look into
the Fig.2 which displays that the longitudinal phase boundary line
$\alpha_{lc}$ concurs with the coercive field line only at the
point $b=0$, i.e., the uniaxial system without transverse field. As
is mentioned above this is the case of no tunneling, and thus we
cannot think of any phase transition. Once $b\neq 0$, i.e., in the
biaxial system, the two lines are separated such that the
second-order transition region becomes larger.

The crossover temperature at the phase boundary is calculated
numerically by using the values on the phase boundary lines
in Fig.3 and the formula $T_c /\tilde{S}A =
(1/2\pi) \sqrt{-v''(x_0)/m}$. Fig.5 shows $T_{zc}/\tilde{S}A$
and $T_{xc}/\tilde{S}A$ as a function of $b_c$ for several values
of $\alpha_x$'s and $\alpha_z$'s, respectively. Since the
$b_c$ at $\alpha_{zc}=0$ decreases as $\alpha_x$ increases (see
Fig.3) the corresponding $T_{zc}(b_c)/\tilde{S}A$ ends at smaller
value of $b_c$ for increasing $\alpha_x$. The same trend can be
found in $T_{xc}(b_c)$.
It is noted that for a
given $b_c$ the crossover temperature at the phase boundary,
$T_{zc}(b_c)/\tilde{S}A$, rises as the transverse field grows, while $T_{xc}(b_c)/
\tilde{S}A$ is lowered as the longitudinal field increases. By fixing the
value of $b_c$ and using Fig.4 we can investigate how the
crossover temperature at the phase boundary varies with the field
parameters. For $b_c = 0.29$ we have shown a 3$d$ plot of
$T_c(\alpha_{xc}, \alpha_{zc})$ in Fig.6. From this picture it can be
easily seen that $T_c$ is lowered with increasing $\alpha_{zc}$,
but rises as $\alpha_{xc}$ increases.

\section{SUMMARY}
\label{sum}
We have studied the phase transition of the escape rate of a
biaxial spin particle with an arbitrarily directed magnetic field.
By using the method of particle mapping we have obtained an
effective particle potential to which the criterion for the
presence of the first-order quantum-classical phase transition
developed in Refs.\onlinecite{blatter,dkpark} has been applied.
From this approach we have computed several phase boundary lines
depending on the transverse anisotropy  and field parameters. In
the field vs transverse anisotropy plot, $\alpha_{xc}(b_c)$ or
$\alpha_{zc}(b_c)$, it is found that the first-order region
decreases as $b_c$ and $\alpha_{x,z}$ increase. In the case of
longitudinal field vs transverse field plot,
$\alpha_{zc}(\alpha_{xc})$, we have observed that compared with the
uniaxial system, both the first- and second-order transition
regions are reduced (with more decrease of the first-order region)
due to the transverse anisotropy. The crossover temperatures at the
phase boundary corresponding to these diagrams have also been
obtained. In the 3$d$ plot of $T_c(\alpha_{xc},\alpha_{zc})$ we
have observed that $T_c$ increases with $\alpha_{xc}$, but
decreases with $\alpha_{zc}$.

Experimentally, the phase diagram of
$\alpha_{zc}(\alpha_{xc})$ can be found in the octanuclear iron
cluster $\rm{Fe}_8$ in which $S=10$ and $b_c = 0.29$. In this case
the intersection with the $\alpha_x$-axis in the phase diagram (see Fig.4) is
given by $\alpha_{xc} = 0.274$ for which the magnetic field is
estimated  to be $H_{xc} = 0.64$T, and for the longitudinal field,
$\alpha_{zc}(0) = 0.81$ in the diagram for which $H_{zc} = 1.9$T,
about three times of $H_{xc}$. The corresponding temperatures are
also estimated to be $T_{zc} = 0.39$K which is the upper limit of
the $T_{zc}(b_c=0.29)$'s for different values of $\alpha_x$'s,
and $T_{xc} = 0.5$K which determines the lower limit of
$T_{xc}(b_c=0.29)$'s for different values of $\alpha_z$'s.

\appendix
\section{COORDINATE DEPENDENT MASS}
In converting the spin Hamiltonian of Eq.(1) to an effective particle
Hamiltonian, if we set the Schr\"{o}dinger-like equation to be
\begin{equation}
\left[- \frac{1}{2m(x)} \frac{d^2}{dx^2} + V(x) \right] \Psi(x) =
E \Psi(x)
\end{equation}
the coordinate dependent mass $m(x)$, effective particle potential
$V(x)$, and the wave function $\Psi(x)$ are derived as following:
\begin{equation}
m(x) = \frac{1}{2A(1+b\sinh^2 x)},
\end{equation}
\begin{eqnarray}
V(x) &=& \frac{A}{4(1+b\sinh^2 x)}\nonumber\\
&\times& [\alpha_z^2 S^2 -4b(S^2 + S -
1/2)- 4\alpha_x S(S+1/2)\cosh x \nonumber\\
&+& b^2 \sinh^4 x - \{b^2 + 4b(S^2+S-1)-\alpha_x^2 S^2 \}\sinh^2 x\nonumber\\
&-& \{4b\alpha_z S(S + 1/2)\cosh x + 2\alpha_x \alpha_z S^2
\}\sinh x ],\nonumber\\
\end{eqnarray}
\begin{eqnarray}
\Psi(x) &=& \left( 1+ b\sinh^2 x
\right)^{\frac{2S-1}{4}}\nonumber\\
&\times& \exp \left[-\frac{\alpha_z S}{2\sqrt{1-b}}
\tanh^{-1}\left(\sqrt{1-b}\tanh x \right) \right]\nonumber\\
&\times& \exp \left[-\frac{\alpha_x S}{2\sqrt{b(1-b)}} \tan^{-1}
\left(\sqrt{\frac{b}{1-b}}\cosh x \right)\right]
\Phi(x)\nonumber\\
\end{eqnarray}
with
\begin{equation}
\Phi(x) = \sum_{\sigma=-S}^S \frac{C_{\sigma}}{\sqrt{(S+\sigma)!
(S - \sigma)!}} \left( \sinh x + \cosh x \right)^{\sigma} .
\end{equation}
In this case, since the mass depends on coordinate the criterion
for the first-order transition becomes more complicated. Following
Ref.\onlinecite{dkpark} the general first-order transition condition
which includes the coordinate-dependent mass is given by
\begin{eqnarray}
[&V'''(x_0 )(g_1 + g_2 /2) + \frac{1}{8}
V''''(x_0 )+ m'(x_0 ) \omega_0^2 g_2 \nonumber\\
&+ m'(x_0 )\omega_0^2 (g_1 +
g_2/2 ) + \frac{1}{4} m''(x_0 )\omega_0^2] <
0, \nonumber\\
\end{eqnarray}
where
\begin{eqnarray}
g_1 (\omega_0 ) &=& - \frac{\omega_0^2 m'(x_0 ) + V'''(x_0 )
}{4V''(x_0 )}, \nonumber\\
g_2 (\omega_0 ) &=& - \frac{2m'(x_0 ) + V'''(x_0 )}{4[4m(x_0 )
\omega_0^2 + V''(x_0 )]},
\end{eqnarray}
and $\omega_0^2$ is the sphaleron oscillation defined as
$\omega_0^2 \ = - V''(x_0 )/m(x_0 )$ from Eq.(15).
Calculating the derivatives of $m(x)$ and $V(x)$ and substituting
these into above condition we have same results as the constant
mass case.

\newpage

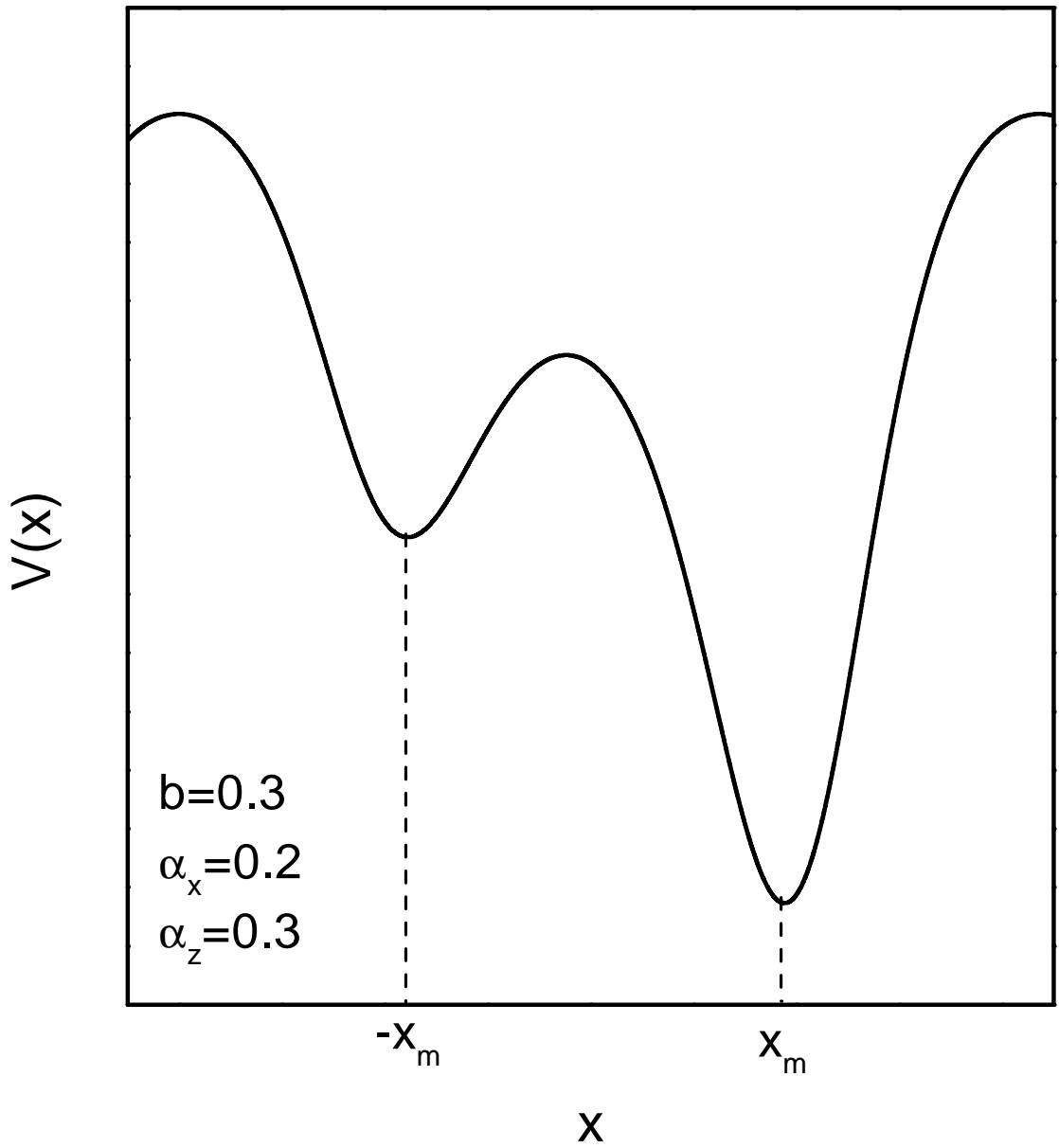
\begin{figure}
\caption{The effective particle potential with $b=0.3$, $\alpha_x = 0.2$,
         $\alpha_z = 0.3$. The local minimum at $x=-x_m$
         corresponds to a metastable state of the spin system
         described in Eq. (1). The inversion process of a spin
         magnetization vector takes place by the escape from the
         local minimum to global minimum at $x=x_m$ along the path
         which passes through the small barrier.
         }
\end{figure}

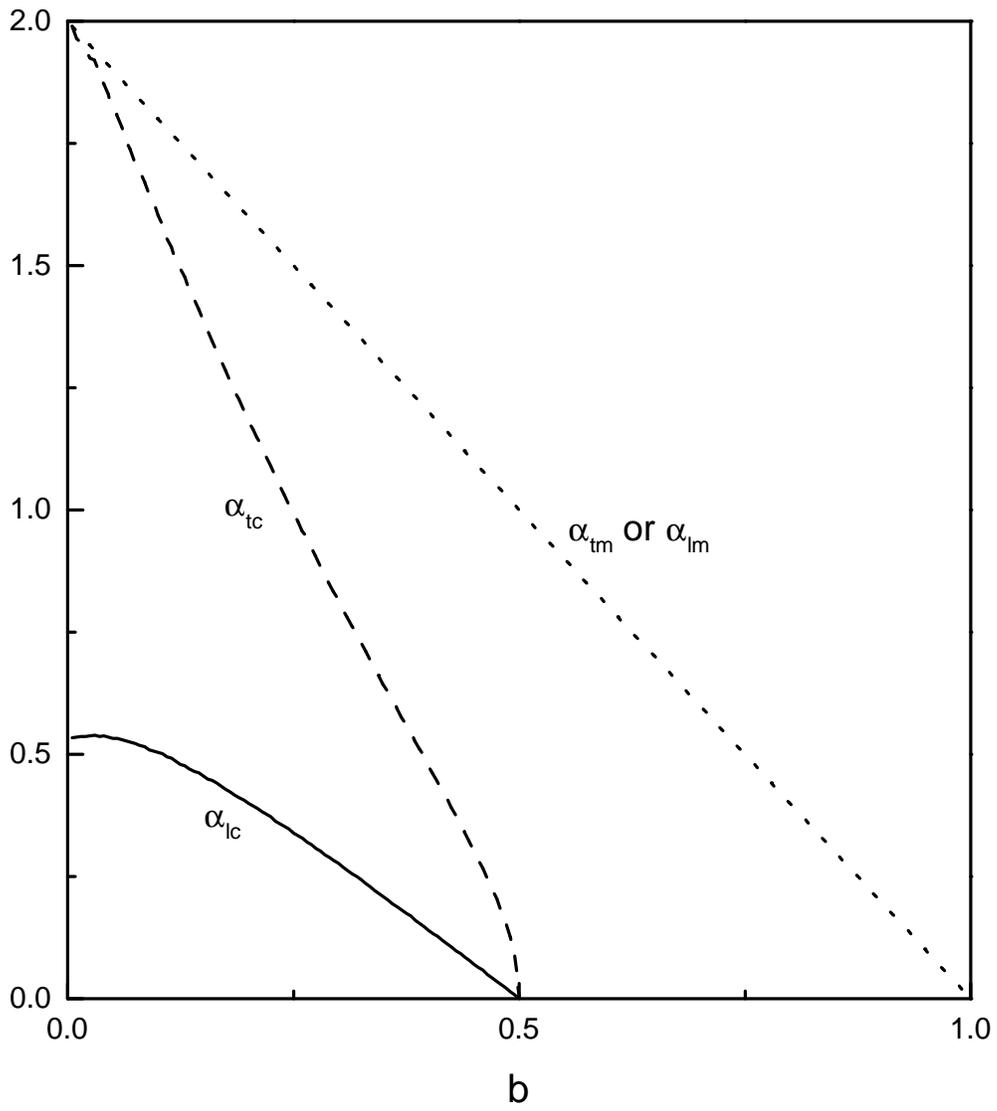
\begin{figure}
\caption{The phase boundary lines $\alpha_{tc}(b_c)$ and
         $\alpha_{lc}(b_c)$, and the coercive field line
         $\alpha_{tm}=\alpha_{lm}\equiv 2(1-b)$.
         }
\end{figure}

\begin{figure}
\caption{The phase boundary lines (a) $\alpha_{zc}(b_c)$'s
         and (b) $\alpha_{xc}(b_c)$'s for several values of
         $\alpha_x$'s and $\alpha_z$'s, respectively.
         }
\end{figure}

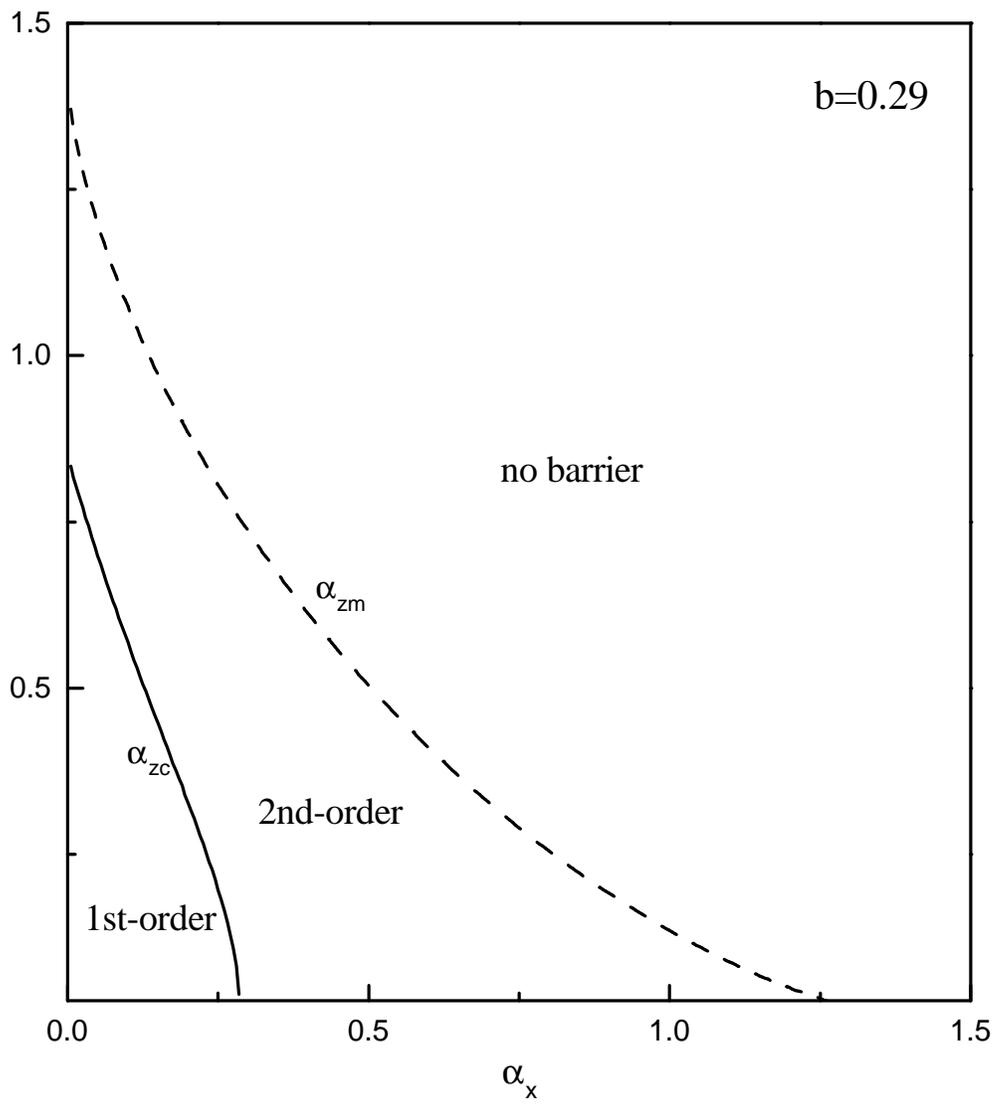
\begin{figure}
\caption{Phase diagram for $b=0.29$ which is the case of
         the iron cluster $\rm{Fe}_8$. The solid line is the
         phase boundary $\alpha_{zc}(\alpha_{xc})$, and the dashed
         line corresponds to the line of metastability condition for
         $b=0.29$.
         }
\end{figure}

\begin{figure}
\caption{Crossover temperatures at the phase boundary as a function of $b_c$:
         (a) $T_{lc}/\tilde{S}A$'s for $\alpha_x = 0$, 0.1, 0.3;
         (b) $T_{tc}/\tilde{S}A$'s for $\alpha_z = 0$, 0.3, 0.7.}
\end{figure}

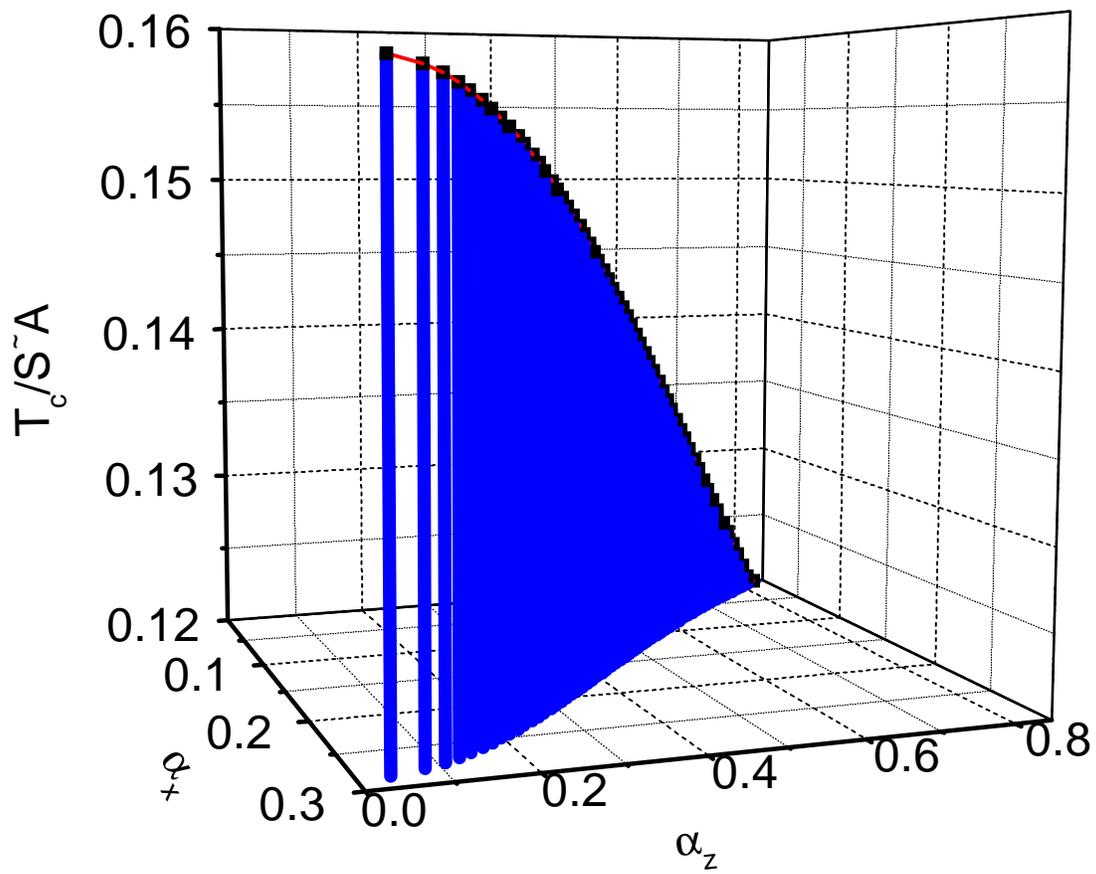
\begin{figure}
\caption{3$d$ plot of the crossover temperature $T_c/\tilde{S}A$
         at the phase boundary as a function of field parameters
         $\alpha_{xc}$ and $\alpha_{zc}$.}
\end{figure}


\begin{thebibliography}{10}
\bibitem{compare1} Comparing with the model in
Ref.\onlinecite{marti} we see $\alpha_{x, z} = 2{\tilde{h}}_{x, z}$ in
the large spin limit.
\bibitem{compare2} It has also been proved in Ref.\onlinecite{lee}
that the small barrier escaping process always dominates even when
the mass depends on the coordinate.
\end{thebibliography}
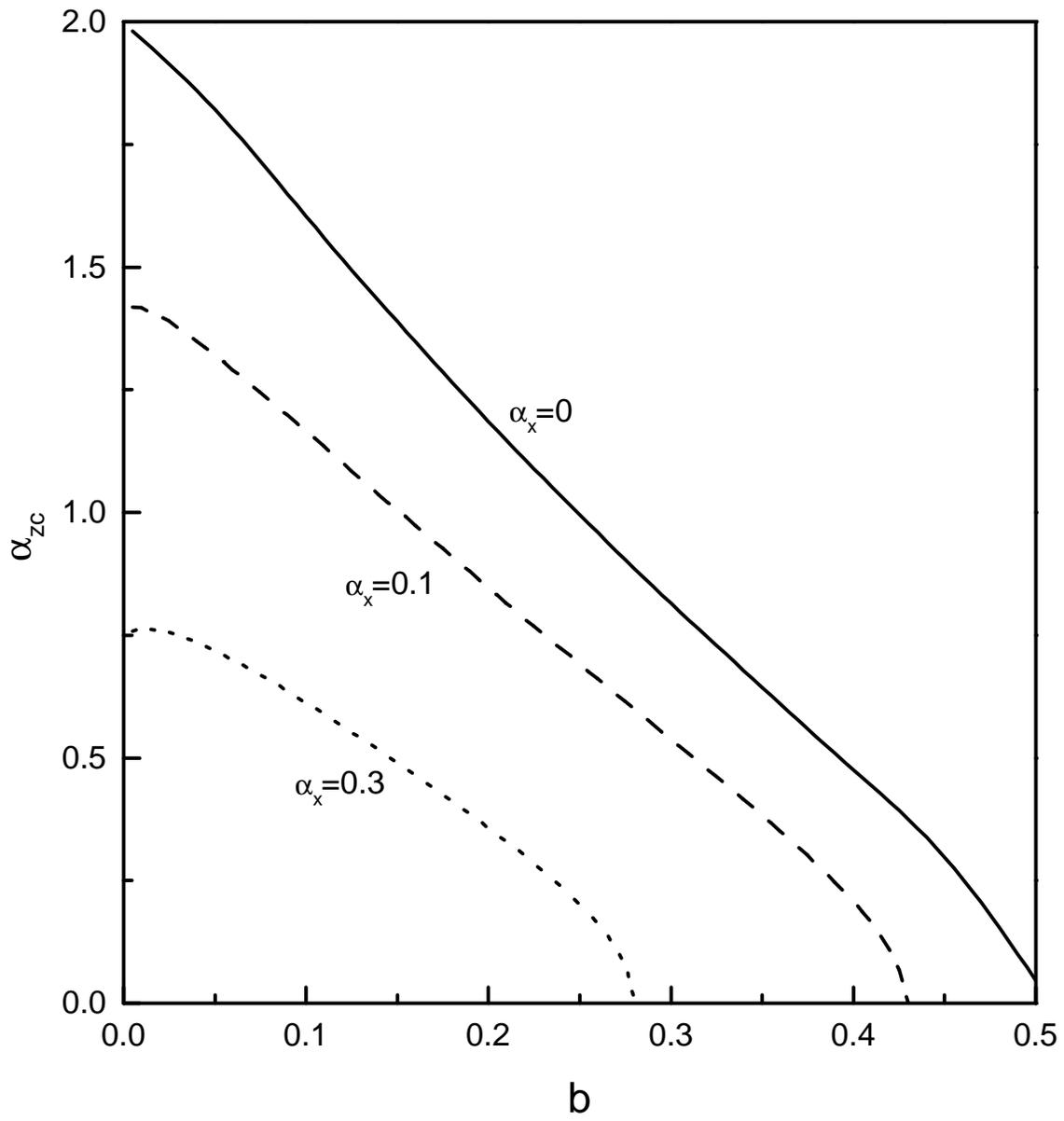
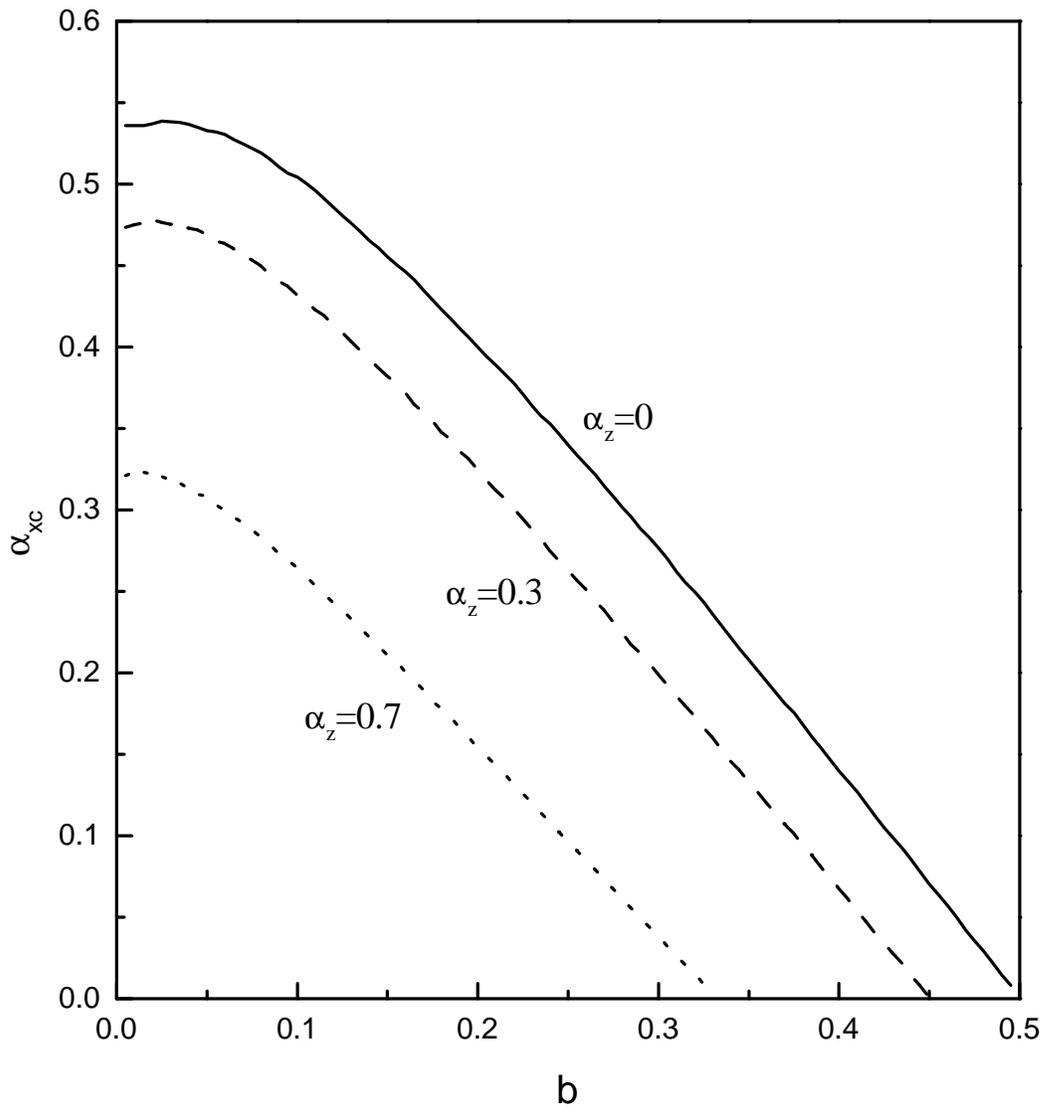
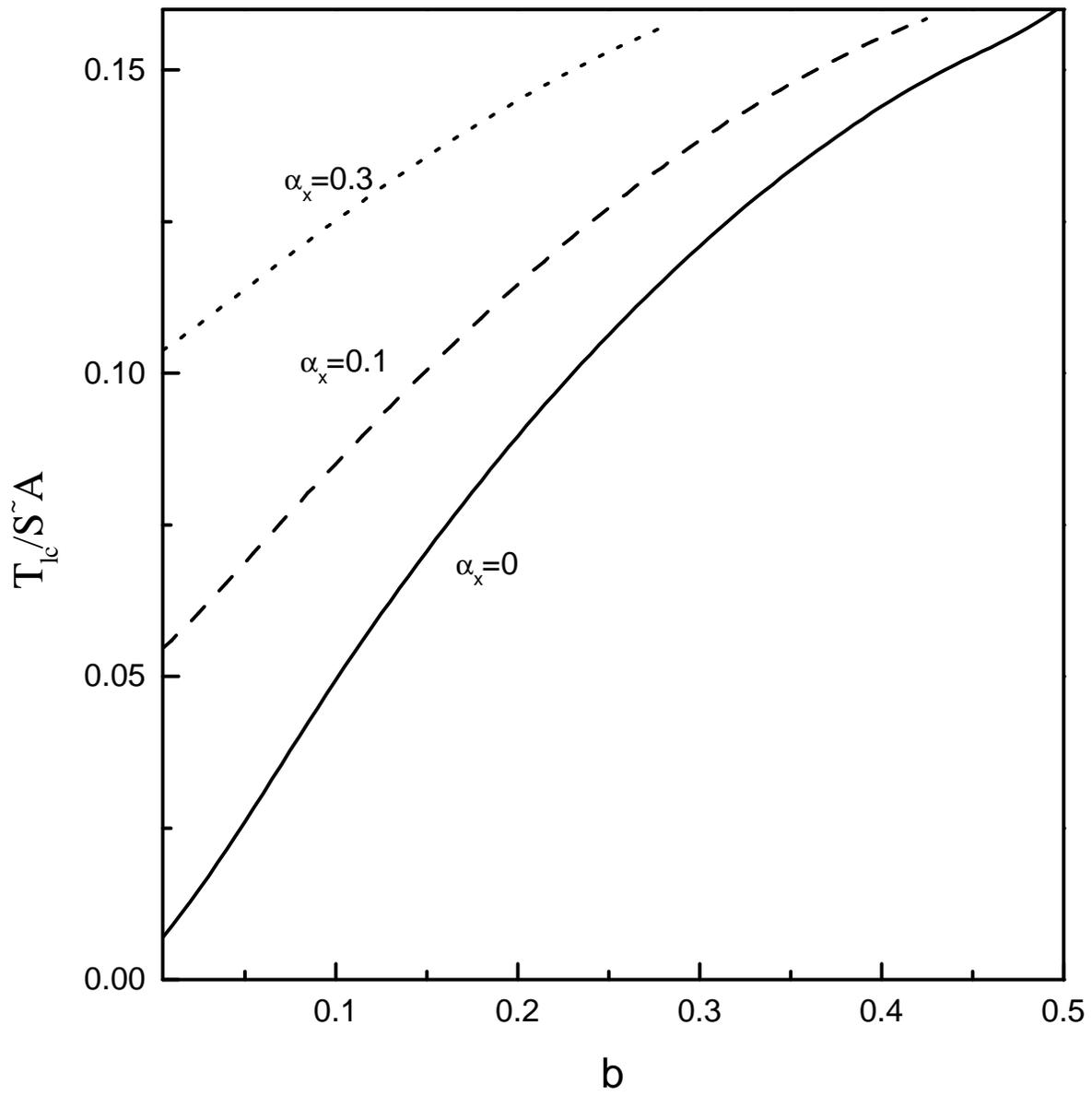
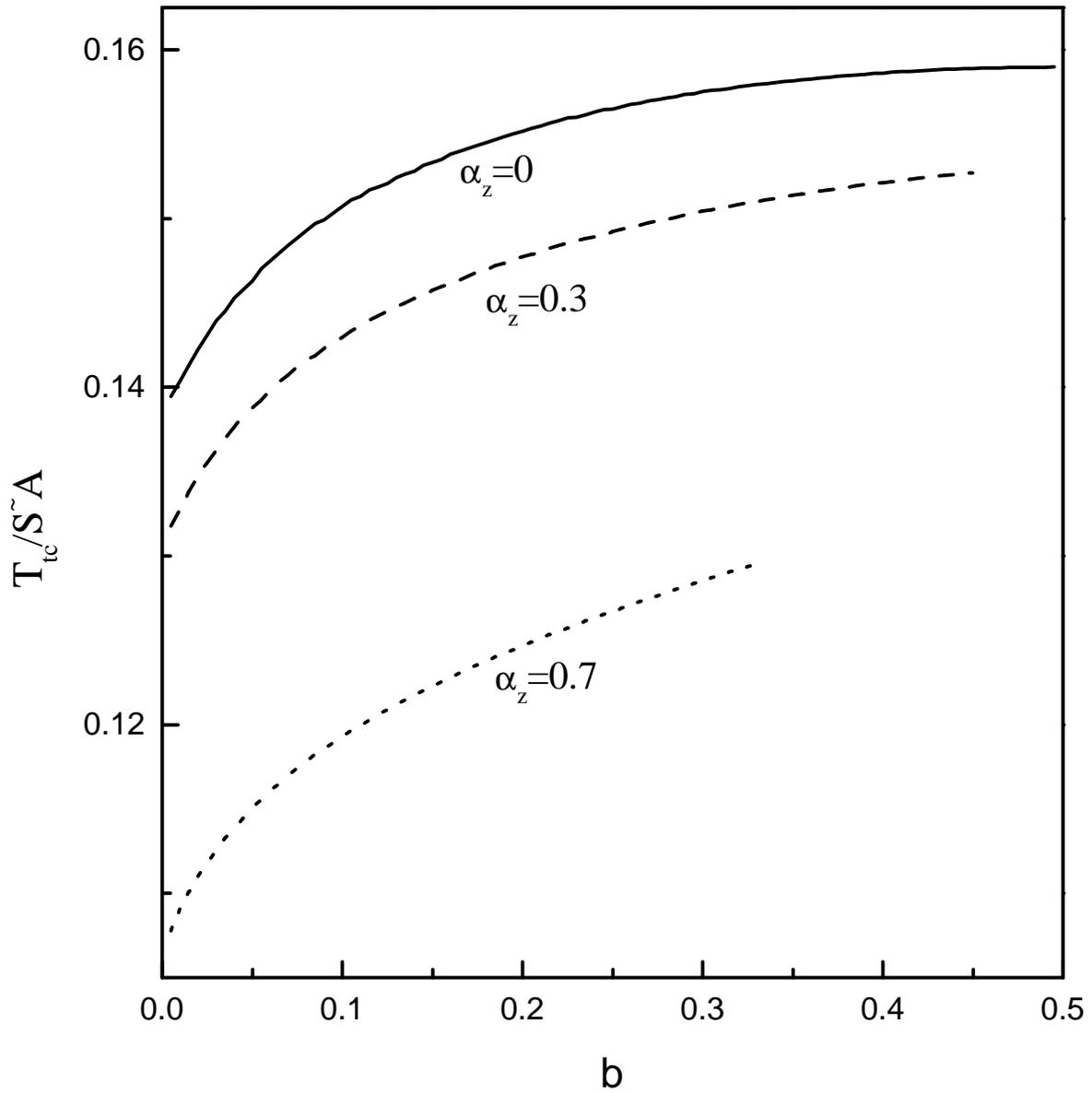
\end{document}